\documentclass[a4]{revtex4}
\usepackage{graphicx}

\begin{document}

\title{Adaptive reflection and focusing of Bose-Einstein condensates}
 \author{A. S. Arnold}
 \email{a.arnold@phys.strath.ac.uk}
 \affiliation{Dept. of Physics, University of Strathclyde, Glasgow G4 0NG, UK}
 \author{C. MacCormick}
 \author{M. G. Boshier}
 \email{m.g.boshier@sussex.ac.uk}
 \affiliation{Sussex Centre for Optical and
Atomic Physics, University of Sussex, Brighton BN1 9QH, UK}

\date{\today}

\begin{abstract}
We report adjustable magnetic `bouncing' and focusing of a dilute $^{87}$Rb Bose gas.
Both the condensate production and manipulation are realised using a particularly
straight-forward apparatus. The bouncing region is comprised of approximately concentric
ellipsoidal magnetic equipotentials with a centre that can be adjusted vertically. We
extend, and discuss the limitations of, simple Thomas-Fermi and Monte-Carlo theoretical
models for the bouncing, which at present find close agreement with the condensate's
evolution. Very strong focusing has been inferred and the observation of atomic
matter-wave diffraction should be possible. Prospects look bright for applications in
matter-wave atom-optics, due to the very smooth nature of the mirror.
\end{abstract}

\maketitle

\section{Introduction}
The realisation of gaseous atomic Bose-Einstein condensation (BEC)
in 1995 \cite{And1} sparked intense theoretical and experimental
interest in this hitherto inaccessible quantum degenerate regime.
Several landmark experiments have followed, and activity in the
field continues to escalate \cite{gsu}. There has been a similar
heightened interest in the area of experimental atom-optics,
following the demonstration of a variety of techniques for
manipulating cold atoms \cite{Hinds}.

As BECs are the ultimate coherent cold atom source, it has become important to develop
high-quality atom-optical elements for manoeuvring them without destroying their
coherence. Potentials typically used for atom-optical manipulation arise from magnetic
and/or optical dipole forces. We will focus in particular on the use of cold atoms in
conjunction with atomic mirrors and lenses. Flat \cite{Kasev1} and curved \cite{Amin}
optical dipole mirrors, as well as flat \cite{Roach,Sid0,Sid1} and curved \cite{Saba1}
magnetic mirrors have all been realised with conventional cold atoms in static
configurations. Dynamic optical dipole mirrors \cite{SteaPh}, flat \cite{Drndic3,Lau1}
and curved \cite{Kadio} dynamic magnetic mirrors, as well as 1D \cite{Marec} and 3D
\cite{mon1,Gor} dynamic magnetic lensing have all been demonstrated. The temporal
variability provided by such mirrors and lenses adds an extra dimension of control.
However, dynamic magnetic mirrors and lenses tend to be softer than their static
counterparts -- i.e. atoms have a longer interaction distance with the mirror potential.
The optical quality of the original static mirrors has been refined \cite{Hinds,Saba1}
and a natural extension is the use of these relatively hard mirrors as atom-optical
elements to focus and image coherent atomic sources such as BECs.

First steps in this direction were made in the experiment of Bongs {\it et al.}
\cite{Bongs}, where the ingredients of a BEC and atomic mirror were successfully combined
with the demonstration of condensate reflection from a flat spatial dipole mirror based
on a rapidly rastered laser beam. Our mirror has the additional advantages that it can be
inelastic, with an adaptable radius of curvature. Here we give a more detailed
description of our earlier work \cite{ArnProc,ArnPR} on the reflection and focusing of a
BEC from this soft curved temporal magnetic mirror. A similar mirror has been used to
reflect, focus and store atom laser beams \cite{til}. The ability to adaptively focus and
launch condensates will enable the realisation of BECs with very high densities, and it
has already enabled us to create extremely cold and coherent collimated atomic clouds.

An important aspect of our mirror design is that all magnetic elements are a relatively
large distance $(d>20\,$mm) from the atoms. Thus, although the magnetic fields are
relatively weak compared to hard atomic mirrors in which the atoms make a close
$(d<100\,\mu$m approach to the magnetic surface, microscopic corrugations in the mirror
potential are drastically reduced because curvature in the magnetic field scales as
$d^{-3}.$ Such large-scale mirrors thus have the advantage that they are extremely
smooth, ensuring fully coherent atomic manipulation. The roughness of our mirror was
measured to be less than $1\,$mrad rms \cite{ArnPR}, where this upper bound was probably
only limited by the sensitivity of the measurement technique, and our atomic mirror is
therefore one of the smoothest in the world \cite{ArnPR,Aspect}.

We begin the discussion with an overview of our `simple' BEC apparatus, followed by
benchmarking experiments that were performed on the condensates. We will then concentrate
on bouncing and focusing BECs: contrasting and discussing the limitations of two
theories, paying particular attention to the theory of experimentally attainable
situations.

\section{`Simple' BEC creation and benchmarking}
\begin{figure}[ht]
 \leavevmode
 \vspace{-3mm}
 \begin{center}
 \mbox{\includegraphics[clip,width=.7 \columnwidth]{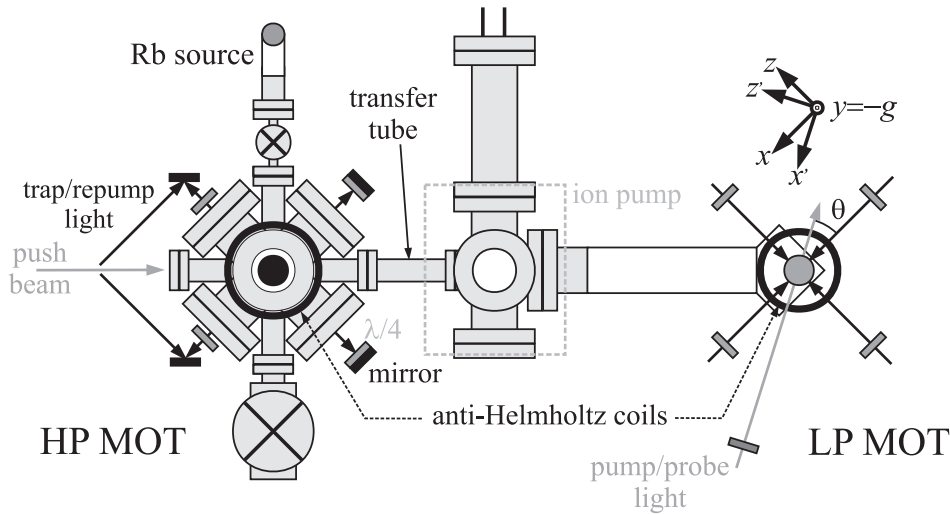}}
 \end{center}
 \vspace{+3mm}
 \caption{The experimental setup, denoting the high (HP) and low (LP) pressure
MOT chambers. Vertical laser beams are not shown. The $20\,$L/s ion pump is all
that is needed to maintain the vacuum.\label{setup}}
 \end{figure}

In the Sussex BEC experiment $N>10^{9}$ $^{87}$Rb atoms can be collected at the low
pressure end of a low-power double magneto-optical trap (MOT) \cite{gib2} system
(Fig.~\ref{setup}). The three extended-cavity diode lasers providing the necessary
$780\,$nm $(5s\,^{2}S_{1/2}\rightarrow 5p\,^{2}P_{3/2})$ MOT light have a very simple,
stable design and can be cheaply manufactured from commercially available components with
minimal machining \cite{arn}. A single diode laser originally provided all trap laser
light for the BEC experiment $(11\,$mW total per MOT in $25\,$mm diameter (FWHM=$20\,$mm)
beams), however two independent lasers are now used for the high pressure and low
pressure MOTs -- increasing the low pressure MOT atom number with a minor cost to
simplicity. For both MOTs the trap laser light is detuned $\Delta=-13\,$MHz (i.e. red) of
the $F=2\rightarrow F'=3$ transition, whilst the repumping light is resonant with the
$F=1\rightarrow F'=2$ line.

Every $80\,$ms the $\approx 10^{7}$ atoms that have accumulated in the high pressure MOT
are transferred to the low pressure MOT simply using the light pressure from a
$150\,\mu$s push pulse of saturated light resonant with the $F=2\rightarrow F'=3$
transition. The shape of the apparatus (Fig.~\ref{setup}) and the relatively small
angular spread of the transferred atoms allow the realisation of a moderate
$(\eta\approx40\%)$ transfer efficiency without requiring a magnetic guiding mechanism.
The low pressure MOT is loaded with $N_{LP}\approx 10^{9}$ atoms in $30\,$s, however
faster loading times could be achieved by inserting an aperture in the transfer tube --
increasing the present $5:1$ MOT pressure ratio.

We find that a dark/compressed MOT phase \cite{ket1} is
unnecessary for BEC production, however this small complication
would no doubt improve the condensate population. The atoms in the
low pressure MOT are cooled to $40\,\mu$K in optical molasses
$(\Delta=-35\,$MHz), then optically pumped into the
$|F,m_{F}\rangle=|2,2\rangle$ state using a circularly-polarised
light pulse resonant with the $F=2\rightarrow F'=2$ transition. A
Ioffe-Pritchard (IP) \cite{Pri} magnetic trap is subsequently
formed around the atoms. The constant magnetic field in the LP MOT
chamber can be altered in $<1\,$ms using three orthogonal pairs of
magnetic coils -- thus providing the necessary field for molasses
and optical pumping as well as positioning the LP MOT centre
relative to the IP trap.

Absorption imaging of the atoms is performed with the optical pumping laser beam. The
imaging lens is an $f=5\,$cm doublet designed to minimise distortion, with a
magnification of $M=0.80(1).$ The short $(8\,\mu$s) pulse of near-resonant imaging light
propagates horizontally at an angle of $\theta \approx \pi/6$ to the $-x$ axis of the
magnetic trap. We refer to the resulting horizontal axis in our images as $-x'$
(Fig.~\ref{setup}). The imaging/optical pumping beam, like the push beam, is derived from
the high pressure MOT trap laser. The intensities of all laser beams derived from the two
trap diode lasers can be altered in times $t<1\,\mu$s via their corresponding
constant-frequency acousto-optic modulators (AOMs). Complete extinction of all laser
light (important during magnetic trapping) is realised using mechanical shutters. As the
free-running rms laser linewidth is $\Delta\nu=300\,$kHz (for times $t\leq 200\,$ms), the
laser frequencies can be precisely adjusted over $\pm 300\,$MHz simply by unlocking the
laser and making small, abrupt changes to the laser current.

The Ioffe-Pritchard magnetic trap coils are formed from
water-cooled $3\,$mm o.d. copper tubing: a 9-turn `cuboidal'
baseball coil with average side dimensions
$45\times45\times55\,$mm$^{3},$ and two three-turn
$64\times101\,$mm$^{2}$ `rectangular' bias coils. The coils'
$10\,$V, $250$A constant-voltage power supply was controlled by
two simple but effective integrator circuits driving independent
water-cooled banks of MOSFETs \cite{mythesis}. Series rms current
noise levels of $\Delta I/I\leq5\times10^{-5}$ were obtained, and
the absence of shunt resistors across both the baseball and bias
coils ensured the maximum level of frequency-insensitive magnetic
noise cancellation when the coils were operated completely in
series \cite{Mewe3}. The magnetic trap's residual bias field of
$10\,$G could be lowered to $0-10\,$G using a pair of 'extra' bias
coils. With a typical compressed magnetic bias field of $1\,$G the
axial and radial trapping frequencies were measured to be
$\nu_{z}=10\,$Hz and $\nu_{r}=223\,$Hz respectively at a magnetic
trap current of $220\,$A.

BEC experiments commonly use two or more large-throughput vacuum pumps, however
due to the relatively small surface area of our apparatus the vacuum is
essentially maintained by a single $20\,$L/s ion pump (Fig.~\ref{setup}). The
lifetime of the magnetic trap, $\tau=70\,$s, is certainly adequate for
obtaining condensates.

After a $32\,$s RF evaporative cooling ramp pure condensates are
reproducibly obtained with more than $10^{5}$ atoms. Two
`bench-marking' experiments were performed on our BEC: the
temperature-dependent ground-state occupation was compared with
the results of Refs. \cite{Ensh1,Mewe2,Han1,AndK1} and the BEC's
ballistic expansion \cite{Ernst2,Steng5} was contrasted with a
Thomas-Fermi \cite{Baym} theoretical model \cite{Cast1}. Our data
yielded good agreement with these earlier results, and more
details of our ground-state occupation experiment will be given
shortly. For our ballistic expansion experiments, absorption
imaging was only valid for condensate expansion times $t>10\,$ms
(cf. \cite{Ernst2}) due to the expected lensing effects
\cite{Andr1}. As in Ref.~\cite{Steng5}, non-destructive imaging
\cite{Andr1} will be used in future to probe the condensate's
behaviour at short release times.

A variety of models have been used by different groups to model
the condensate and non-condensate fractions of cold atomic clouds.
Three density distributions commonly used in BEC data analysis
will be summarised here:
\begin{itemize}
  \vspace{-2mm}
  \item{the Gaussian distribution \\ $n_{G}({\bf r})=
  \frac{N}{(2\pi)^{3/2} \prod_{i=1}^{3}{\sigma_{r_i}}}\exp\!\left[
    -\sum_{i=1}^{3}{{r_i}^{2}/(2{\sigma_{r_i}}^2})\right]$}
  \vspace{-2mm}
  \item{the `Bose' distribution \\ $n_{B}({\bf r})=
      \frac{N}{g_{3}(1)(2 \pi)^{3/2}\prod_{i=1}^{3}{\sigma_{r_i}}}\,
      g_{3/2}\!\left[\exp\!\left(
    -\sum_{i=1}^{3}{{r_i}^{2}/(2{\sigma_{r_i}}^2)}\right)\right]$ \\
    where $g_{n}$ is the polylogarithm function $g_n(z)=\sum_{k=1}^{\infty}{z^k/k^n}.$}
  \vspace{-2mm}
  \item{the Thomas-Fermi distribution \\
  $n_{TF}({\bf r})=\frac{15 N}{8 \pi \prod_{i=1}^{3}{A_{r_i}}}
    \max\!\left[1-\sum_{i=1}^{3}{{r_i}^{2}/A_{r_i}^2},0\right].$}
  \vspace{-2mm}
 \end{itemize}

The density distribution obtained from absorption imaging is
integrated along the line of sight of the probe laser beam. By
using the co-ordinate transformation into the $x'y'z'$ (imaging)
frame
 \begin{equation}
 \left(\begin{array}{c}
  x'\\y'\\z'
 \end{array}\right)=
\left(\begin{array}{ccc}
  \cos\theta&0&\sin\theta\\0&1&0\\-\sin\theta&0&\cos\theta
 \end{array}\right)
 \left(\begin{array}{c}
  x\\y\\z
 \end{array}\right)
 \end{equation}
and integrating along the absorption imaging beam axis $(x'),$ one
obtains the equivalent `two-dimensional' density for each of these
distributions:
   \begin{equation}
   n_{G2}(y',z')=\frac{N}{2 \pi \sigma_{y'} \sigma_{z'}}\exp\!\left[
    -(\frac{y'^2}{2{\sigma_{y'}}^2}+\frac{z'^2}{2{\sigma_{z'}}^2})\right],
   \label{G2D}
   \end{equation}
   \begin{equation}
   n_{B2}(y',z')=\frac{N}{g_3(1) 2 \pi\sigma_{y'}\sigma_{z'}}\,
      g_{2}\!\left[\exp\!\left(
    -(\frac{y'^2}{2{\sigma_{y'}}^2}+\frac{z'^2}{2{\sigma_{z'}}^2})\right)\right],
   \label{B2D}
   \end{equation}
   \begin{equation}
   n_{TF2}(y',z')=\frac{5N}{2\pi A_{y'} A_{z'}}
              \max\left[1-\frac{y'^2}{{A_{y'}}^2}-\frac{z'^2}{{A_{z'}}^2},0\right]^{3/2},
   \label{TF2D}
   \end{equation}
where the width parameters $\sigma_{y'},$ $\sigma_{z'},$ and the
maximum TF radii $A_{y'},$ $A_{z'}$ have the same angular
dependence, i.e.
  \begin{equation} \sigma_{y'}=\sigma_{y},\;\;\sigma_{z'}=\sqrt{{\sigma_{z}}^{2}\cos^{2}\theta+{\sigma_{x}}^{2}\sin^{2}\theta},
   \;\;A_{y'}=A_y,\;\;A_{z'}=\sqrt{{A_z}^{2}\cos^{2}\theta+{A_x}^{2}\sin^{2}\theta}.
  \label{TFwidthdash}
  \end{equation}
In the Sussex BEC experiment the viewing angle of the camera is
$\theta\approx \pi/6$ because the probe laser optical access to
the vacuum cell is constrained by the IP magnetic trap coils. In
order to obtain a viewing angle along one of the magnetic trap
axes it would be necessary to directly overlay the probe laser
beam with the MOT laser beams. This could be done in future by
placing polarising beamsplitters in front of the MOT quarter-wave
plates -- thus allowing light to be coupled into the MOT beam
path, then coupled out again after passing through the vacuum
cell.

\begin{figure}[ht]
 \leavevmode
 \vspace{-3mm}
 \begin{center}
 \mbox{\includegraphics[clip,width=.7 \columnwidth]{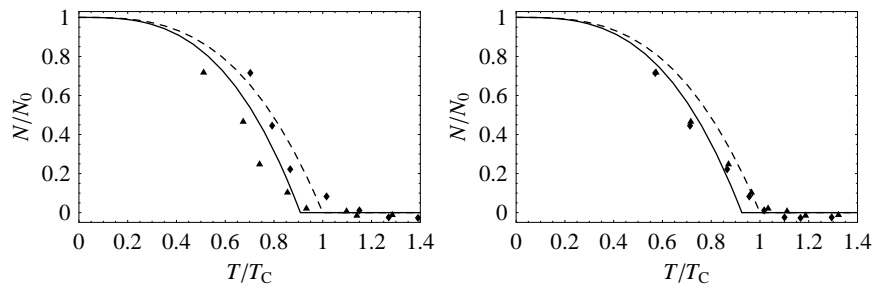}}
 \end{center}
 \vspace{+3mm}
 \caption{Relative condensate population as a function of temperature. The data
was obtained by fitting bi-modal distributions to the average of
five imaged thermal/condensate clouds after a $23\,$ms ballistic
expansion. Data points of increasing temperature correspond to
final RF evaporative cooling frequencies of $710,
715,...,760\,$kHz. Thomas-Fermi distributions were fit to the BEC
clouds. In the left image both the temperature and number of atoms
in the thermal cloud was obtained from a fitted Gaussian
(triangles) or Bose (diamonds) distribution. In the right image
the atomic populations were obtained in the same way, however the
temperature was obtained by fitting a Gaussian to the wings of the
distribution. \label{NTgraf}}
 \end{figure}

The different models used by previous groups to measure the condensate ground state
occupation used either a bi-modal Gaussian distribution \cite{AndK1,Han1}, a
`Bose'/Thomas-Fermi distribution for the thermal/condensate populations \cite{Mewe2}, or
a bimodal Gaussian distribution with a separate Gaussian fit to the `wings' of the
distribution to measure the temperature \cite{Ensh1}. We found that, for our trap
parameters and within experimental uncertainties, the atomic distributions were fit
equally well with a Gaussian/Thomas-Fermi or Bose/Thomas-Fermi distribution. We found
that bi-modal Gaussian fits were consistently worse compared to these two models. This
was expected due to the high condensate density $(n_0(t=0)\approx2\times
10^{14}\,\mbox{cm}^3)$ in our magnetic trap, which ensured the applicability of the
Thomas-Fermi approximation. Fig.~\ref{NTgraf} illustrates the Gaussian/Thomas-Fermi and
Bose/Thomas-Fermi fits. In the left graph the temperature is derived directly from the
Gaussian or Bose distribution fitted to the thermal atoms, whereas in the right graph the
temperature is determined by fitting a Gaussian to the `wings' of the distribution (like
in Ref.~\cite{Ensh1}).

\section{Bouncing and focusing condensates}

Atoms moving adiabatically in a magnetic field experience a potential,
 $U({\bf r})=-${\boldmath$\mu$}$\cdot {\bf B}= g_{F} m_{F} \mu_{B} B,$
that depends on the magnitude of the magnetic field, $B,$ as well as the strength and
sign of their atomic state's magnetic moment in the direction of ${\bf B}$ $(g_{F} m_{F}
\mu_{B})$. Initial magnetic focusing experiments with neutral atoms \cite{mon1,Marec,Gor}
have involved cylindrically symmetric systems comprised of one or two circular coils.
Judicious application of pulses of magnetic potential with positive axial curvature (and
negative radial curvature) followed by a magnetic potential pulse with negative axial
curvature allows one to three-dimensionally focus a cold cloud of atoms which passes
through the coils \cite{mon1,Gor}.

Ideally, one would like to focus the atoms with a single magnetic
pulse as the dynamics are much simpler. The BEC experiment already
relies on a magnetic potential with positive curvature in all
three spatial dimensions -- i.e. the magnetic trap -- this trap is
therefore ideal for 3D atom focusing, and for large enough fields
it can also be used as a mirror. Another advantage of using the
same coils for BEC formation and magnetic bouncing/focusing is
that one does not need to precisely position the respective
centres of two different coil systems. A simpler, separate mirror
consisting of two coils in a quadrupole configuration can be built
\cite{Kadio}, however along the axis of the coils the magnetic
field magnitude varies linearly, resulting in only 2D focusing.

A final pair of magnetic coils was used to generate a vertical bias field
 ${\bf B}_{c}=\{0,B_{c},0\}$
whilst the atoms are `bouncing' on the IP potential. The time-dependent magnetic
field to second-order is thus:
 \begin{eqnarray}
 {\bf B}&=&B_{0}\{0,0,1\}+B_{1}\{x,y_{c}-y,0\} \nonumber \\
          & &+B_{2}/2\{-x z,-y z,z^2-(x^2+y^2)/2\}
  \label{IPfield}
 \end{eqnarray}
where the IP trap's magnetic bias, gradient and curvature
$(B_{0},$ $B_{1},$ $B_{2}$ respectively) are proportional to the
series current in the trap coils, and the centre of the magnetic
trap is shifted vertically upward by the variable amount
$y_{c}=B_{c}/B_{1}.$ The magnitude of the magnetic field is
parabolic in the axial $(z)$ direction, but hyperbolic in the
co-ordinate $r=\sqrt{x^2+(y-y_{c})^2}$ and thus linear for radii
$r>\sqrt{2}B_0/B_1.$

\begin{figure}[ht]
 \leavevmode
 \vspace{-3mm}
 \begin{center}
 \mbox{\includegraphics[clip,width=.7 \columnwidth]{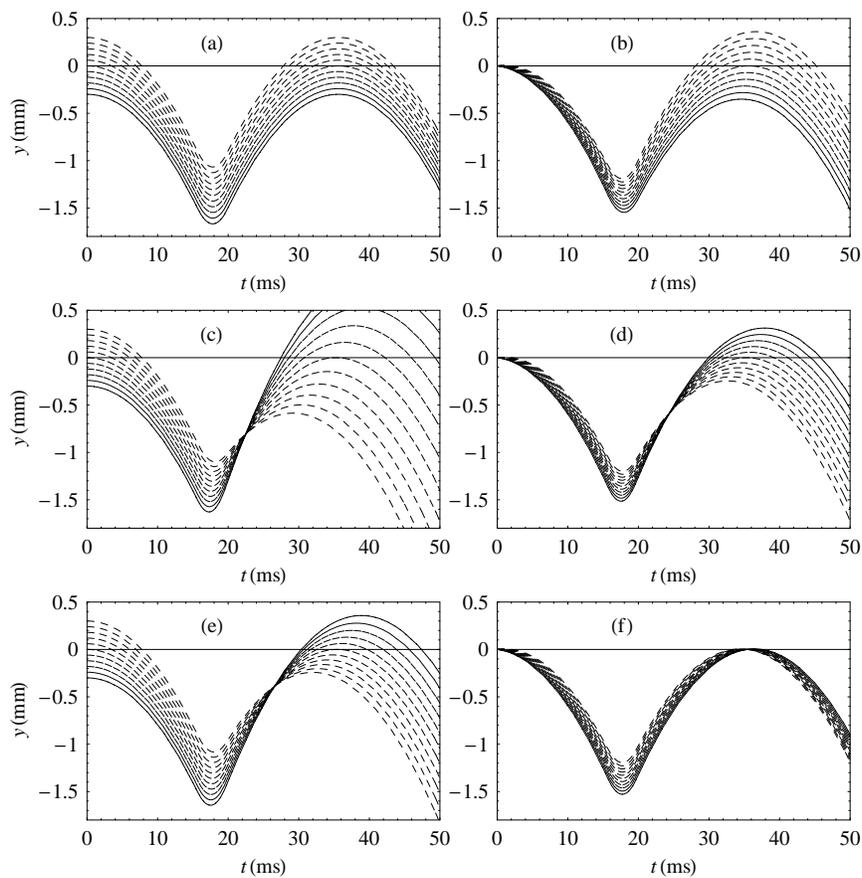}}
 \end{center}
 \vspace{+3mm}
 \caption{One dimensional bouncing on a pulsed potential. The left images ((a), (c), (e)) show the trajectories
of atoms with zero initial velocity and a range of heights. The right images ((b), (d),
(f)) show the trajectories of atoms with zero initial position and a range of velocities.
Images (a) and (b) show reflection from a linear potential. Images (c) and (d) show
reflection from a parabolic potential if the drop height $h$ is equal to the radius of
curvature of the mirror $R.$ Images (e) and (f) show reflection from a parabolic
potential if the drop height is $h=R/2.$\label{bouncetute}}
 \end{figure}

In the experiment, a condensate is released at time $t=0$ and
expands ballistically until $t=15\,$ms when both the IP trap (at
$\approx70\%$ of the $220\,$A maximum current) and a constant
vertical bias field $(B_{c}=0-400\,$G) are pulsed on for $5\,$ms.
After $30\,$ms an identical magnetic field pulse allows the
realisation of a second `bounce'. During the `bounce' pulses the
atoms experience a magnetic field magnitude that is weakly
confining (parabolic) in the $z$ direction, strongly confining
(with strength dependent on ${\bf B}_{c})$ in the $x$ direction
and almost linear in the $y$ direction (see Fig.~1 of
Ref.~\cite{ArnPR}). Linear spatial dependence of the magnetic
field magnitude results simply in a change of the centre-of-mass
motion (bouncing), whereas parabolic dependence yields
focusing/defocusing of the atomic cloud (Fig.~\ref{bouncetute}).

During reflection the COM motion changes slightly as the vertical magnetic bias is
altered because its field has a small amount of curvature. Horizontal motion arises
partly due to a tilt in the IP trap with respect to gravity, and also due to the fact
that the magnetic potentials have an increasingly large slope in the $z$ direction as $y$
decreases. This $yz$ term in the magnetic field magnitude also causes the cloud to begin
rotation in the $x=0$ plane during the bounce.

Atoms bouncing in a gravity-cavity are unstable when they are dropped from a height
greater than half the radius of curvature of the bouncing surface $(h>R_{r_i}/2)$
\cite{cctifo}.

We have experimentally studied situations where the drop height compared to the mirror
radius of curvature in the $x$ direction is $h=R_{x}$ and $h=R_{x}/5.4$ respectively,
with the results illustrated in Fig.~2 of Ref.~\cite{ArnPR}. Due to our imaging angle,
the instability of the condensate in the $x$ direction is not seen in the $z'\approx z
\cos(\pi/6)+x \sin(\pi/6)$ direction until after the second bounce.

We now turn to the theoretical analysis of the bouncing experiments. It has been shown
\cite{Cast1} that a BEC, in the Thomas-Fermi regime and experiencing a time-dependent
parabolic potential
 $U({\bf r},t)=\frac{m}{2}\sum_{i=1}^{3}{\omega_{{r}_{i}}(t)^2{r_{i}}^2},$
has an atomic spatial distribution which obeys the simple scaling law:
 \begin{eqnarray}
 n({\bf r},0)= \mbox{max}\left[\frac{n_0}{\prod_{i=1}^{3}{\lambda_{r_{i}}}}
 \left(1-\sum_{i=1}^{3}{\frac{{r_{i}}^2}{A_{r_{i}}\lambda_{r_{i}}}}\right),0\right]
 \label{TF1}
 \end{eqnarray}
where $\lambda_{r_{i}}(t)$ is the solution of
 \begin{equation}
 \frac{d^2\lambda_{r_{i}}(t)}{dt^2}+\omega_{r_{i}}(t)^2\lambda_{r_{i}}(t)-
     \frac{\omega_{r_{i}}(0)^2}{\lambda_{r_{i}}(t)\;\,\prod_{j=1}^{3}{\lambda_{r_{j}}(t)}}=0,
 \label{TF2}
 \end{equation}
 $\lambda_{r_{i}}(0)=1$ and the initial BEC maximal radii are $A_{r_{i}}.$
We note that if the parabolic potential has a time-varying centre
 ${\bf r}_c(t),$ i.e. if
 $U({\bf r},t)=\frac{m}{2}\sum_{i=1}^{3}{\omega_{{r}_{i}}({\bf r}_c(t))^2(r_{i}-{r_{i}}_c(t))^2},$ then the
{\it same} scaling law applies for the BEC if one subtracts out the
centre-of-mass (COM) motion ${\bf r}_0(t)$ and uses the magnetic curvatures
acting at the COM.

The second theoretical model is a Monte-Carlo simulation. Initial positions for
$j=1,2,...,M$ atoms distributed on a spherical Thomas-Fermi distribution, $n(r)\propto
\max[1-r^2,0],$ can be obtained by inverting the equations arising from the cumulative
density function ---
\[\rho_{\theta\phi r}\propto\int{d\theta}\int{\sin\phi\,d\phi}\int{r^2\,n(r)\,dr}=
 \rho_\theta\,\rho_\phi\,\rho_r=\theta\,\cos\phi\,(r^3/3-r^5/5)\] --- i.e. by uniformly choosing
$\theta_j\in[0,2\pi],$ $\rho_{r_j}\in[0,2/15],$ and $\rho_{\phi_j}\in[-1,1]$ and by
inverting $\rho_{r_j}$ and $\rho_{\phi_j}$ to obtain $r_j$ and
$\phi_j=\arccos(\rho_{\phi_j}).$

The initial cartesian positions of the atoms ${\bf r}_j={\bf r}_1,...,{\bf r}_M$ in a
cylindrically symmetric trap $(A_x=A_y=A_r)$ are then given by
\[{\bf r}_j=r_j (A_r\sin(\phi_j)\cos(\theta_j),A_r\sin(\phi_j)\sin(\theta_j),A_z\cos(\phi_j)),\]
where $A_r=\frac{1}{\pi\nu_r}\sqrt{\frac{\mu}{2m}},$
$A_z=\frac{1}{\pi\nu_z}\sqrt{\frac{\mu}{2m}}$ and $\mu(N)=(2 \pi)^{6/5} m^{1/5}(15
\hbar^2 N a {\nu_r}^2 \nu_z)^{2/5}/2$ 
 is the chemical potential. The BEC's spatial distribution has a $t_{d}=15\,$ms expansion
in which the co-ordinates ${\bf r}_j$ are scaled by Eq.~\ref{TF2}, with corresponding
velocities
 ${\bf v}_{j}=\{\frac{d\lambda_x}{dt}(t) x_{j},
  \frac{d\lambda_y}{dt}(t)y_{j},\frac{d\lambda_z}{dt}(t)z_{j}\}.$ For times $t>15\,$ms,
when the density of the condensate and thus the effect of interatomic repulsion has
dropped dramatically, the atoms are allowed to follow their classical ballistic
trajectories.

Our magnetic coils were simplified to line segments, and the Biot-Savart law was then
used to generate the magnetic field and its Taylor expansion. We could thus ensure that
in our case no important third- or higher-order terms are neglected in the Monte-Carlo
model if one uses the magnetic field magnitude from approximation Eq.~\ref{IPfield}. For
the Thomas-Fermi model we use Eq.~\ref{IPfield} to derive the local magnetic field
curvatures -- which vary slightly with the centre-of-mass position of the BEC during the
bounce pulses.

In the Monte-Carlo simulation the condensate maintains an approximately Thomas-Fermi
distribution (Eq.~\ref{TF2D}) in the $yz'$ imaging plane throughout the bouncing
sequence. However, due to the $yz$ term in the magnetic potential, the distribution is
rotated in the $yz$ plane by an angle $\phi_{yz}$ (or $\phi_{yz'}$ in the $yz'$ plane).
In order to find the radii, and tilt angle $\phi_{yz'},$ of the Monte-Carlo distribution,
the cartesian co-ordinates of the $M$ Monte-Carlo `atoms' are projected into the $yz'$
plane with co-ordinates $(y_i,z_i').$ It can be shown that given the moments of the cloud
in the $yz'$ plane
\[\langle z' \rangle=\frac{1}{M}\sum_{i=1}^{M}{z'_i},\;\;\;\langle y
\rangle=\frac{1}{M}\sum_{i=1}^{M}{y_i},\;\;\;\langle yz'
\rangle=\frac{1}{M}\sum_{i=1}^{M}{y_i\,z'_i},\;\;\;\langle z'^2
\rangle=\frac{1}{M}\sum_{i=1}^{M}{{z'_i}^2},\;\;\;\langle y^2
\rangle=\frac{1}{M}\sum_{i=1}^{M}{{y_i}^2}\] one obtains the rotation angle of the cloud:
 \[\phi_{yz'}=\frac{1}{2}\arctan\left(\frac{\langle yz' \rangle}
 {{z'_{rms}}^2-{y_{rms}}^2}\right),\]
 where $z'_{rms}=\sqrt{\langle z'^2 \rangle-\langle z' \rangle^2}$ and $y_{rms}=\sqrt{\langle y^2 \rangle-\langle y
 \rangle^2}$ are the rms `radii' of the cloud in the non-rotated $yz'$ frame.
Additionally, in the $yz'$ frame rotated by $\phi_{yz'},$ the true rms radii of the
condensate are given by
\[Z_{rms}=\sqrt{{z'_{rms}}^2 \cos^2\phi_{yz'}+{y_{rms}}^2 \sin^2\phi_{yz'}+\langle yz' \rangle\sin(2\phi_{yz'})}\]
\[Y_{rms}=\sqrt{{z'_{rms}}^2 \sin^2\phi_{yz'}+{y_{rms}}^2 \cos^2\phi_{yz'}-\langle
yz'\rangle\sin(2\phi_{yz'})}\]
 and one can show that these rms radii can be converted to Thomas-Fermi distribution maximum radii
 by multiplying by a factor of $\sqrt{7}.$

\begin{figure}[ht]
 \leavevmode
 \begin{center}
 \mbox{\includegraphics[clip,width=.6\columnwidth]{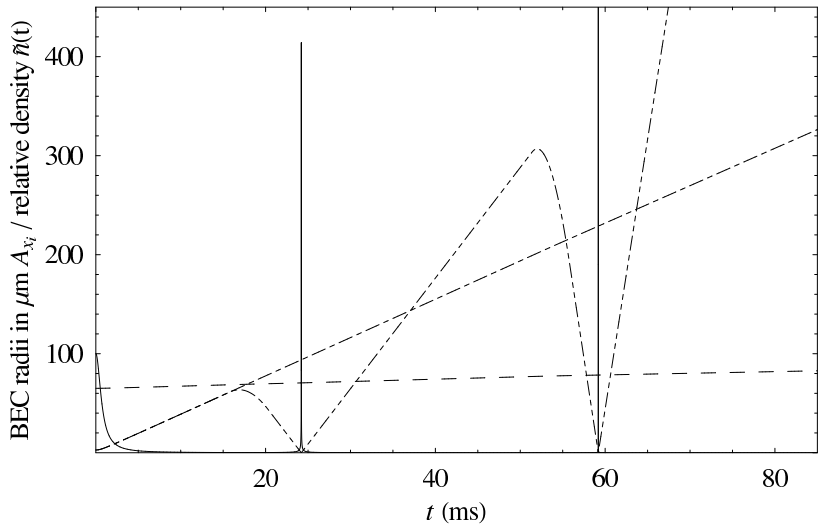}}
 \end{center}
 \caption{The time-evolution of the BEC's $x,$ $y$ and $z$ radii ($A_x$ dash-dot-dotted, $A_y$ dash-dotted
 and $A_z$ dashed) in microns obtained using the Thomas-Fermi model. Also of interest is the scaled relative density
 $\tilde{n}(t)=100\,n_0(t)/n_0(0)\propto 1/(A_x A_y A_z),$ which, following a slow decay, has two sharp peaks when $A_x$
 briefly comes to a tight focus. The initial peak density in the condensate is $n_0(0)=2\times
 10^{14}\,\rm{cm}^{-3},$ and at a time $t_{\rm{f}}=24.2\,\rm{ms}$ (shortly after the condensate bounces for the first
 time), the peak density is four times larger for a duration of $\approx10\,\mu\rm{s}$. Subsequent foci of $A_x$ are even tighter, however they depend strongly
 on the harmonicity of the mirror.
 \label{Theory2}}
 \end{figure}

The theoretical models work well \cite{ArnPR} in describing the absorption images taken
from our current viewing angle, but they also show that even more interesting behaviour
should be seen in the $xy$ plane. When $B_c=0\,$G (i.e. $h=R_x$) then after the first
bounce the condensate comes to a tight focus in the $x$ direction. The Thomas-Fermi model
(Fig.~\ref{Theory2}), with its inclusion of atomic interactions but neglect of anharmonic
spatial magnetic fields, predicts a very tight focus with an $x$ cloud maximum radius of
$A_x=18\,$nm. The de Broglie wavelength of the BEC at the focus, $38\,$nm, can be
obtained from its velocity. Thus, if a suitably harmonic mirror is used and the focused
matter wave has a transverse width smaller than the matter wavelength, one may expect to
see atomic diffraction during the focus .

We now compare the Thomas Fermi model with three versions of the Monte-Carlo model. The
Monte-Carlo models use: the full magnetic potential originating from Eq.~\ref{IPfield}
(MC), the local harmonic approximation to the magnetic potential at the centre of mass
(MC$_{\rm{H}}$), and the harmonic MC$_{\rm{H}}$ model with the inclusion of the effects
of the $x^2y$ local potential aberration term (MC$_{{\rm H}x^2y}$). We note that the
MC$_{\rm{H}}$ model and Thomas-Fermi model are identical during the initial condensate
expansion $(t<15\,\rm{ms}),$ and it is only during focusing - if the density is large
enough for interatomic repulsion to be important - that the models may differ.

\begin{figure}[ht]
 \leavevmode
 \begin{center}
 \mbox{\includegraphics[clip,width=.7\columnwidth]{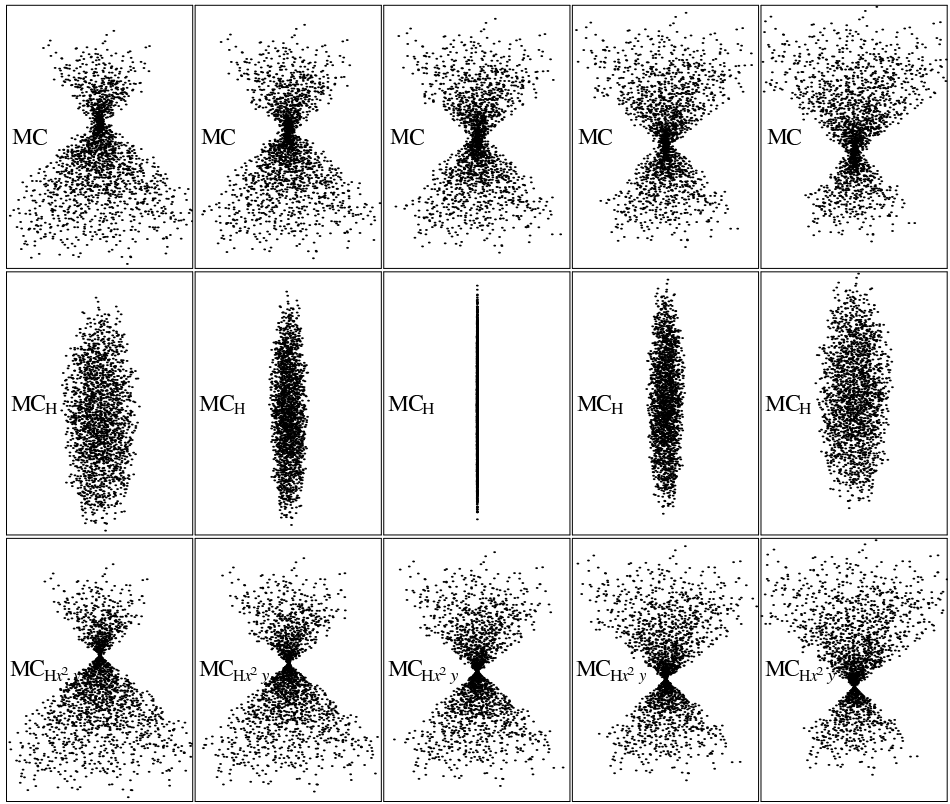}}
 \end{center}
 \caption{A theoretical comparison of the temporal foci of the atomic distribution in the $xy$ plane subsequent
to the first bounce. Three different Monte-Carlo models are used (as described in the
text): MC, MC$_{\rm{H}}$ and MC$_{{\rm H}x^2y}$ models are illustrated in the upper,
middle and lower rows of the figure, respectively. Each image has $x\times y$ (horizontal
$\times$ vertical) widths of $4\,\mu\rm{m}\times 200\,\mu\rm{m},$ and the images
correspond to times (left to right)
 $t=t_{\rm f}-30\,\mu{\rm s},t_{\rm f}-20\,\mu{\rm s},...,t_{\rm f}+10\,\mu{\rm s}$ where
$t_{\rm{f}}=24.2\,\rm{ms}$ is the time at which the focus occurs in the Thomas-Fermi
model. \label{xyfoc}}
 \end{figure}

In Fig.~\ref{xyfoc} we look at the behaviour of the foci in the $xy$ plane as predicted
by the various Monte Carlo models. We find that the MC model has a maximum density of
$n_0=9(2)\times10^{14}\,\rm{cm}^3$ in a `bow-tie' shaped focus which varies location with
time. The MC$_{\rm{H}}$ model has an extremely tight elliptical focus, leading to a peak
density of $n_0=24(2)\times10^{14}\,\rm{cm}^3$ -- this density is more than ten times
higher than the Thomas-Fermi model in which interatomic repulsion limits the density at
the condensate focus. The MC$_{{\rm H}x^2y}$ model has a similar (but crisper) focal
shape and a similar maximum density $(n_0=10(2)\times10^{14}\,\rm{cm}^3)$ to the MC model
-- this illustrates that the `bow-tie' focus is mainly generated by the $x^2y$ terms in
the magnetic potential from Eq.~\ref{IPfield}. The approximate (FWHM) temporal duration
of the above Monte Carlo density maxima during the focus is $30,\,2,\,30\,\mu\rm{s}$ for
the MC, MC$_{\rm{H}},$ and MC$_{{\rm H}x^2y}$ models respectively. This can be compared
to the $10\,\mu\rm{s}$ duration of the focus seen in the Thomas-Fermi model. In addition,
the foci of the Monte Carlo models occur approximately $10\,\mu\rm{s}$ before the
Thomas-Fermi focus, as interatomic repulsion delays the focal time.

The Thomas-Fermi model thus predicts very tight foci in the $xy$ plane, however the
anharmonicities of our present mirror mean that our Monte-Carlo model will be better at
predicting the shape of the $xy$ focus. The Monte-Carlo models' neglection of the
interatomic repulsion which is included in the Thomas-Fermi model, also appears to be
invalid, as the MC$_{\rm{H}}$ model differs dramatically from the Thomas-Fermi model at
the focus. During the focus very high densities are obtained, but only for a very short
duration compared to the time over which interatomic repulsion is important during the
initial condensate expansion $(\approx 2\,\rm{ms}$ as can be seen in Fig.~\ref{Theory2}).
Thus although we might expect the MC models to break down during the focus, the expansion
trajectory of the condensate after the focus may well be valid.

When we begin probing such extremely small length scales in the $xy$ plane (nanometres --
rather than the $>1\,\mu\rm{m}$ diffraction-limited resolutions of present BEC imaging
systems), both of the simple theories described here - the Thomas-Fermi and Monte-Carlo
models - will therefore cease to be valid. These experiments will need to be compared to
calculations using the more complicated time-dependent 3D Gross-Pitaevskii equation,
however it is worth noting that the Thomas-Fermi model should be adequate if a
sufficiently harmonic magnetic mirror is employed, and the Monte Carlo model may well
suffice for describing condensate behaviour prior to and immediately after focusing if
the tightness of the focus does not induce atomic matter-wave diffraction.

In conclusion, we have created a relatively simple and robust BEC apparatus. With this
apparatus we have demonstrated adjustable focusing and reflection of a BEC from a soft,
extremely smooth, and adaptive temporal magnetic mirror. If highly parabolic magnetic
mirrors of this type are used, this technology should enable one to focus atoms down to
nanoscales. In order to investigate such focusing, we plan to image the condensates in
the $xy$ plane in future, with the hope of observing atomic matter-wave diffraction.
Highly parabolic mirrors can also create extremely cold, collimated atomic beams for use
in atomic clocks and we have already demonstrated a rudimentary matter-wave `beam
expander' in this vein \cite{ArnPR}.
 \vspace{2mm}

This work was supported by the UK EPSRC and the University of Sussex. ASA gratefully
acknowledges funding from the Commonwealth Scholarship Commission and the Royal Society
of Edinburgh.

\end{document}